\documentclass[journal=langd5,manuscript=article]{achemso}
\usepackage{amsmath}
\usepackage{graphicx}
\usepackage{color}
\usepackage{ulem}

\newcommand{\revision}[1]{{\textcolor{black}{#1}}}

\author{Patrick Grosfils}
\affiliation[Microgravity, Universit\'{e} Libre de Bruxelles]
{Microgravity Research Center, Chimie Physique E.P. CP 165/62, Universit\'{e} Libre de Bruxelles, Av.F.D.Roosevelt 50, 1050 Brussels, Belgium.}
\alsoaffiliation[CENOLI, Universit\'{e} Libre de Bruxelles]
{Center for Nonlinear Phenomena and Complex Systems CP 231, Universit\'{e} Libre de Bruxelles, Blvd. du Triomphe, 1050 Brussels, Belgium}
\author{James F. Lutsko}
\email{jlutsko@ulb.ac.be}
\affiliation[Universit\'{e} Libre de Bruxelles]
{Center for Nonlinear Phenomena and Complex Systems CP 231, Universit\'{e} Libre de Bruxelles, Blvd. du Triomphe, 1050 Brussels, Belgium}

\title{The low-density/high-density liquid phase transition for model
globular proteins}

\begin{document}

\begin{abstract}
The effect of molecule size (excluded volume) and the range of interaction on the surface tension\revision{, phase diagram and nucleation properties} of a model globular protein is investigated using a combinations of  Monte Carlo simulations and finite temperature classical Density Functional Theory calculations. We use a parametrized potential that can vary smoothly from the standard Lennard-Jones interaction characteristic of simple fluids, to the ten Wolde-Frenkel model for the effective interaction of globular proteins in solution. We find that the large excluded volume characteristic of large macromolecules such as proteins is the dominant effect in determining the liquid-vapor surface tension and nucleation properties. The variation of the range of the potential only appears important in the case of small excluded volumes such as for simple fluids. \revision{The DFT calculations are then used to study homogeneous nucleation of the high-density phase from the low-density phase including the nucleation barriers, nucleation pathways and the rate. It is found that the nucleation barriers are typically only a few $k_{B}T$ and that the nucleation rates substantially higher than would be predicted by Classical Nucleation Theory.}
\end{abstract}

\date{\today }

\section{Introduction}

One of the most important problems in biophysics is the characterization of
the structure of proteins. It is well-known that the main impediment to the
determination of protein structure is the difficulty with which good quality
protein crystals can be produced. This has led to a large body of work
focused on the understanding the details of protein nucleation. In recent
years, it has become apparent as a result of simulation, theoretical and
experimental studies that nucleation in general, and protein nucleation in
particular, is strongly affected by the presence of intermediate, metastable
states\cite{VekilovCGDReview2004,tWF,NicolisPhysica,GuntonProtein,LutskoNicolis}. 
This raises the possibility that by understanding the mechanism by
which intermediate states affect nucleation, the quality of the final result
can be better controlled.

\revision{The practical importance of investigating the role of intermediate metastable states lies in the fact that the effective interactions of proteins in solution depend on their environment. Protein molecules interact via Coulombic forces mediated by the ionic solution in which they are dissolved. The effective interaction between two protein molecules therefore depends on the properties of the solution, particularly the salt used to create the solution and its pH. This is the reason for the well-known fact that some salts are more effective than others in precipitating protein crystallization (the Hofmeister effect)\cite{Hof}. Detailed confirmation of the connection between the properties of the solution and the effective intra-protein interactions has come from both computer simulation\cite{Jungwirth,Netz} and from theoretical studies of the phase diagram of proteins in solution\cite{Gunton}. One goal of the present work is to investigate what aspect of the effective interactions is most relevant in controlling the nucleation of the metastable phase.}

Assuming that the effects of the solvent can, at first approximation, be
entirely accounted for by an effective interaction potential between protein
molecules, the problem of protein nucleation can be viewed as analogous to
the nucleation of a solid from a dilute gas. In this way, Wilson observed
that favorable conditions for protein nucleation are correlated with the
behavior of the osmotic virial coefficient\cite{GeorgeWilson}. Rosenbaum, 
et al. showed that
the phase diagram of a large class of globular proteins can be mapped onto
that of simple liquids with an interaction potential that depends on the
ionic strength of the solvent \cite{Rosenbaum1, Rosenbaum2}. One particular characteristic of proteins is
that these models involve a hard-core repulsion and an attractive tail with
the range of the attraction being quite small compared to the size of the
hard core. It is known that as the range of the attraction becomes smaller,
the critical point of the liquid-vapor transition is suppressed relative to
the triple point until for sufficiently short-ranged potentials, the
liquid-vapor transition becomes metastable with respect to the vapor-solid
transition \cite{LutskoNicolis}. It is in this circumstance that the meta-stable liquid phase is
thought to play an important role in protein nucleation.

Despite the abundance of evidence for the role of the intermediate state
from both simulation and experiment, there is still no really convincing
theoretical description based on first-principles. Lutsko and Nicolis showed that
transitions from vapor to metastable liquid to solid appeared advantageous
relative to the direct vapor-solid transition based on the bulk free energy surface\cite{LutskoPRL}, but that work neglected the
effect of interfaces and, in particular, surface tension. The goal of the
present work is to take a step towards filling in this gap by characterizing
the liquid-vapor interfacial surface tension and the liquid-vapor transition in a model globular protein consisting of molecules interacting with the ten Wolde-Frenkel potential\cite{tWF}. However, because of
the metastable nature of the transition, such a characterization based on
simulation is difficult. We found it impossible to stabilize the liquid-vapor interface for the model protein
due to the strong tendency towards crystallization. We
have therefore had to use an indirect approach consisting of a combination
of theory and simulation. First, the ten Wolde-Frenkel model potential for
globular proteins is generalized so that it depends on three independent
parameters that allow it to be deformed continuously from a Lennard-Jones
potential, i.e. a simple fluid, to the hard-core+tail ten Wolde-Frenkel (tWF)
potential. We have performed simulations covering part
of the range from simple fluid to protein and compare this to Density
Functional Theory calculations to show that the DFT remains quantitatively
accurate as the hard-core radius is increased from zero, in the LJ
potential, to a typical value in the tWF potential and as the range of
attraction decreases. Whereas simulation becomes infeasible when the range becomes too
short, the calculations are easily performed and the preceding agreement
gives some confidence in the result. The DFT is then used to calculate the
nucleation barrier using recently developed energy surface techniques. We find, somewhat surprisingly, that the increase in
the hard-core radius is much more important than the 
decrease in the range of the potential and is mostly responsible for a 
dramatic drop in surface
tension. This observation is particularly relevant in the case of
proteins since the properties of the effective interaction can be varied by changing controllable parameters, such as the PH of the solution, so that a primary goal is to determine those conditions most favorable to homogeneous nucleation. Our results indicate that, so far as the low-density/high-density part of the transition is concerned, varying the range of the potential has little effect.

In the next Section, our model potential is defined and the Monte Carlo
simulations described and the DFT calculations are also sketched. In Section III, we describe the comparison between theory and simulation as the potential is varied from Lennard-Jones towards the model protein interaction and show quantitative agreement between theory and simulation. We then present the DFT results for the surface tension and nucleation barrier and nucleation rate as calculated from DFT and compare to classical nucleation theory. The paper
concludes with a discussion of our results.

\section{Theoretical  Methods}

\subsection{Simulation}

\begin{figure*}[tbp]
\includegraphics[angle=0,scale=0.6]{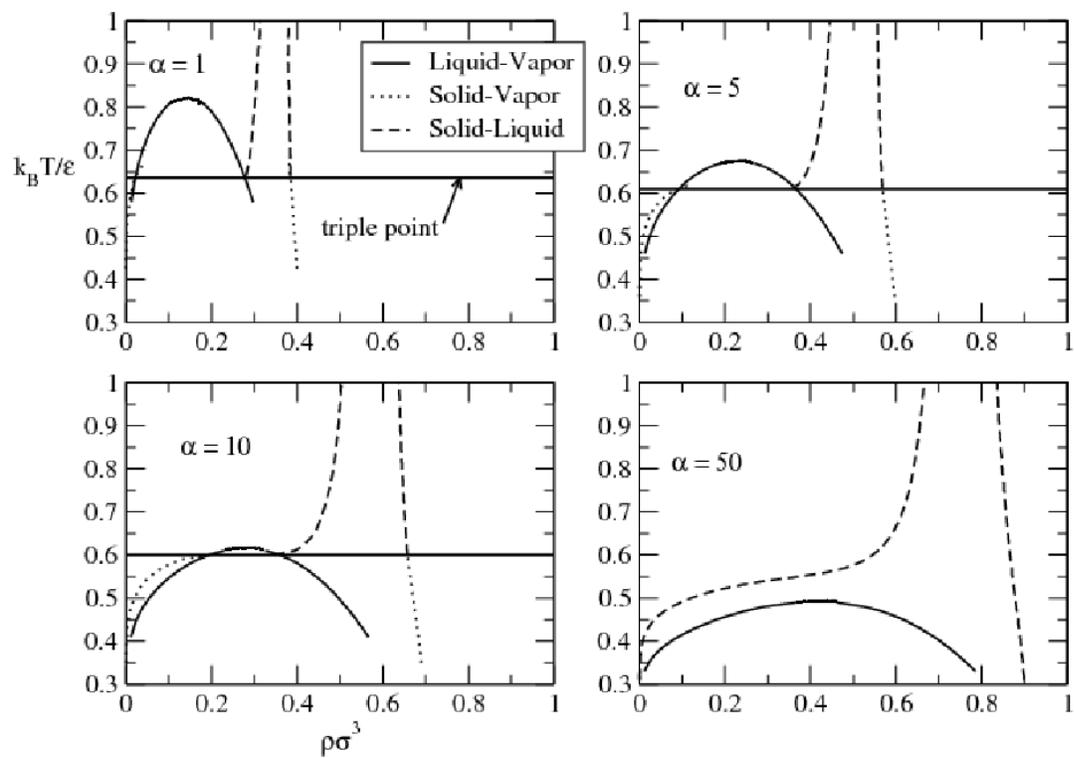}
\caption{Calculated phase diagrams as a function of $\alpha$ for $\delta = 1$ showing
that the liquid is metastable  $\protect\alpha \ge 10$. From Lutsko and Nicolis\cite{LutskoNicolis}.}
\label{fig1a}
\end{figure*}

Because of the complexity of the constituent particles and of the solvent-induced interaction
globular proteins are often modeled, in a first approximation, by an effective interaction potential.
In their study of the phase behavior of globular proteins in solution ten Wolde and Frenkel proposed the following effective potential \cite{tWF}
\begin{equation}\label{a1}
V(r)\,=\,\frac{4\,\epsilon}{\alpha^2}
\left(\,
\left( \frac{1}{(\frac{r}{\sigma})^{2} -1 }  \right)^{6}-\,\alpha\,
\left( \frac{1}{(\frac{r}{\sigma})^{2} -1 }  \right)^{3}
\right).
\end{equation}
This is a Lennard-Jones potential modified to capture the essential features that characterize
the interaction of proteins in solution. The potential includes a hard-sphere repulsion 
at $r=\sigma$ that accounts for the excluded volume, i.e. the size, of the protein molecule. The parameter $\alpha$
controls the range of solvent induced effective attraction in that the minimum of the potential well occurs at $r=\sigma \sqrt{1+\left( \frac{2}{\alpha} \right)^{1/3}}$ while the minimum of the potential is $-\epsilon$, independent of the value of the range. 
The range of the potential decreases when $\alpha$ increases and for $\alpha=50$ this model potential
reproduces the phase behavior of globular protein solutions and with in
particular the well known fact that for sufficiently short range attractions the liquid 
phase becomes thermodynamically unstable\cite{tWF, LutskoNicolis}, as illustrated in \ref{fig1a}. For the potential (\ref{a1}) this behavior is expected 
to be found when $\alpha \simeq 10$ in accordance with the generally accepted 
criteria according to which the range of the attraction should be $\sim 25\%$ of the range of the 
repulsive part to have a metastable liquid-gas coexistence\cite{LutskoNicolis}.

Metastable systems are notoriously difficult to study numerically because the liquid
state spontaneously decays to the more stable crystal phase. In the present situation, 
the value $\alpha=50$ puts the liquid-gas coexistence curve deep inside the unstable region which 
makes the liquid phase difficult to maintain. A possible solution to this problem is to
use constraints that restrict the configurational space of the system thereby preventing the
transition to the solid phase. There are different ways to constrain the system \cite{nie:154505}. 
In the restricted Monte Carlo method \cite{Corti19942717} the density is constrained to be below a 
limited value which thus suppresses the dense phase. While this approach is well suited to the study of the supersaturated vapor
phase it is not appropriate for studying supercooled liquids because the densities of the liquid and 
solid phases are too close. Another possibility would be to limit the number of neighbors per particle.
This approach has been applied with success to a Lennard-Jones potential.
However there is a body of evidence that besides its effect on the stability of liquids, the range of
the attractive potential impacts also on the structure of the liquid \cite{Doye}.
This fact could be of special importance in the present case because the 
structure of the liquid is atypical due to the large
excluded volume which limits the number of interactions per particle. 
More generally the shortcomings of restricted ensemble methods is the elimination of 
many configurations.

In this work no constraints were imposed on the system. Stable liquid-gas equilibrium were
obtained starting from a stable two-phase state and modifying step by step the temperature and the 
parameter $\alpha$ until the desired conditions are met. Unfortunately for the desired case of $\alpha=50$ this 
was not possible because there is no stable configuration to start from. We therefore generalized the potential as
\begin{equation}\label{a2}
V(r)\,=\,\frac{4\,\epsilon}{\alpha^2}
\left(\,
\left( \frac{1}{(\frac{r}{\sigma})^{2} -\delta^2 }  \right)^{6}-\,\alpha\,
\left( \frac{1}{(\frac{r}{\sigma})^{2} -\delta^2}  \right)^{3}
\right).
\end{equation}
\revision{with a hard core at $r < \delta \sigma$}, so that, by varying the dimensionless parameters $\alpha$ and $\delta$, it is possible to smoothly go from a  \revision{simple} fluid, described by the Lennard-Jones interaction with $\alpha = 1$ and $\delta = 0$ to the model protein interaction with $\delta = 1$ and $\alpha = 50$. This allows
to reach a metastable liquid-gas coexistence from a stable Lennard-Jones
system by varying in small steps the temperature and the parameters $\alpha$ and $\delta$.

We simulated an ensemble of $N=1885$ particles interacting via the potential (\ref{a2}) with a 
standard Metropolis Monte-Carlo algorithm (MC-NVT)\cite{FrenkelSmit}. The potential is truncated at $r_c\,=\,2.8$ but not
shifted. The particles are contained in a volume $V$ with dimensions
$L_x\,=\,L_y\,=9\sigma$, and $L_z\,=\,108\sigma$ and periodic boundary conditions are imposed in 
all directions. To avoid the problem of stability of the liquid phase, the system is equilibrated 
during $5\times 10^5$ Monte-Carlo cycles (one cycle $=\,N$ updates)
at given ($T$, $\alpha$, $\delta$) starting from a stable initial configuration at 
($T+\Delta T$, $\alpha+\Delta \alpha$, $\delta+\Delta \delta$). 
The configuration of the system consists 
in a liquid slab of thickness $\Delta z\simeq 27\sigma$ surrounded along the z-direction by two gas slabs.
In this way the stability of the liquid phase was maintained during the $10^6$ cycles used to measure 
the density profile and the surface tension. 

Surface tension can be measured by different methods. In a recent paper we adopted the Bennett 
method as this method appears to be precise \cite{PGLutsko}. However we found it difficult
to implement here for the following reason. In the Bennett's method the calculation of the surface tension follows from the definition 
\begin{equation}\label{a3}
\gamma\,=\,\left(\frac{\partial F}{\partial A}\right)_{N,V,T}
\end{equation}
where $F$ is the free energy and $A=L_x\times L_y$ is the area of each liquid-vapor interface. 
In its implementation the method requires that one compares the energies $E_i$ ($i=0\,,1$)
of two configurations with different liquid-vapor interface areas $A_0$ and $A_1$. 
The configuration with interface area $A_1$ is obtained from the previous one by rescaling the 
positions of the particles \cite{Salomons}\cite{FrenkelSmit}:
\begin{eqnarray}\label{a4}
x'&=&x\,(A_1/A_0)^\frac{1}{2}\,,\nonumber\\
y'&=&y\,(A_1/A_0)^\frac{1}{2}\,,\nonumber\\
z'&=&z\,(A_0/A_1)\,. 
\end{eqnarray}
Because this perturbation is done at constant volume, the system is expanded along one direction 
and compressed along the transverse direction. Due to the isotropy of the fluid phase this
transformation has only a negligible effect on the energy of homogeneous systems because the energy 
change created by the displacement along the compression direction is compensated by an energy change
along the expansion direction. In the presence of an interface 
the symmetry is lost and there is a net energy difference between the two configurations 
which is localized at the liquid-vapor interface. While the perturbation (\ref{a4}) is usually an efficient way 
to probe the interfacial free-energy we found it to be unsuitable here as the surface tension of the 
protein model is particularly small so that large perturbations are necessary to have a measurable free energy difference. Unfortunately large perturbations are inefficient when applied to systems whose potential contains an excluded volume because
the compression step creates configurations with overlaps between particles. This not only gives infinite energy variation
but these variations are often located inside the liquid phase and not at the interface. To solve this problem we
implemented the interface wandering 
approach which allows a precise evaluation free-energy difference with small area perturbations\cite{macdowell:061609}.
In this method the system is perturbed at each
Monte Carlo cycle according to (\ref{a4}). Contrary to the Bennett's method where the new configuration is
only tested, in the interface wandering method the new area, chosen at random in an interval $[A_{\rm min}\,,A_{\rm max}]$, is effectively accepted with probability 
\begin{equation}\label{a5}
P_{\rm accepted}\,=\,{\rm min}\left(1\,,\,e^{-\beta\,(E_1-E_0)}\right).
\end{equation}
As a result of this acceptance ratio, the area of the interface evolves and is distributed between 
$A_{\rm min}$ and $A_{\rm max}$ according to a distribution $f$, which is related to the
free energy
\begin{equation}\label{a6}
F\,=\,-k_BT\ln f\,
\end{equation}
Because in this method many areas are sampled, instead of one in
the Bennett's method, a curve fitting can be performed on the interface area distribution function
which allows a precise determination of the surface tension.

\subsection{DFT calculations}

The properties of inhomogeneous systems were calculated using Density
Functional Theory. According to DFT, the free energy in the grand canonical
ensemble, the grand potential $\Omega $, is a functional of the local
density, $\rho \left( \mathbf{r}\right) $, and the applied external field, $%
\phi \left( \mathbf{r}\right) $, of the form%
\begin{equation}
\Omega \left[ \rho \right] =F\left[ \rho \right] +\int \left( \phi \left( 
\mathbf{r}\right) -\mu \right) \rho \left( \mathbf{r}\right) d\mathbf{r}
\end{equation}%
where $\mu $ is the chemical potential,  $F$ and $\Omega $ both depend on
temperature and the functional $F\left[ \rho \right] $ does not depend on the
field\cite{EvansDFT, HansenMcdonald}. The equilibrium density distribution is determined by minimizing $%
\Omega \left[ \rho \right] $. In a uniform system, the local density is  a
constant, $\rho \left( \mathbf{r}\right) =\overline{\rho }$ and $F\left[
\rho \right] \rightarrow F\left( \overline{\rho }\right) $ is the Helmholtz
free energy.

In our calculations, $F\left[ \rho \right] $ is approximated  using the
Modified-Core van der Waals DFT model\cite{Lutsko_JCP_2008}. This is a generalization of the simplest hard-sphere plus mean-field tail model which gives quantitatively accurate descriptions of fluid structure\cite{Lutsko_JCP_2008}, surface tension\cite{Lutsko_JCP_2008,PGLutsko} and nucleation properties\cite{Lutsko_JCP_2008_2}. Here, we require quantitative accuracy in the DFT calculations as we wish to make direct comparison to simulation. In this model, the free
energy is expressed as a sum of three contributions:%
\begin{equation}
F\left[ \rho \right] =F_{HS}\left( \left[ \rho \right] ;d_{HS}\right)
+F_{core}\left( \left[ \rho \right] ;d_{HS}\right) +F_{tail}\left( \left[
\rho \right] ;d_{HS}\right) .
\end{equation}%
The first term on the right is the hard-sphere contribution which is treated
using the Fundamental Measure Theory functional. The second term is the
''core correction''\ which is similar in form to the hard-sphere term, but
which modifies the hard-sphere contribution so that the free energy in the
bulk phase reproduces a given equation of state\cite{Lutsko_JCP_2008}. The last term is the tail
contribution and has the simple mean field form%
\begin{equation}
F_{tail}\left( \left[ \rho \right] ;d_{HS}\right) =\frac{1}{2}\int \Theta
\left( r_{12}-d_{HS}\right) \rho \left( \mathbf{r}_{1}\right) \rho \left( 
\mathbf{r}_{2}\right) v\left( r_{12}\right) d\mathbf{r}_{1}d\mathbf{r}_{2}.
\end{equation}%
The reference hard-sphere diameter, $d_{HS}$, is calculated using the
Barker-Henderson\cite{BarkerHend,HansenMcdonald} expression,%
\begin{equation}
d_{HS}=\int_{0}^{r_{0}}\left( 1-e^{-\beta V\left( r\right) }\right) dr
\end{equation}%
where $r_{0}$ is the distance at which the potential vanishes, $V\left(
r_{0}\right) =0$. 

The model requires as input the bulk equation of state. For this, we use the
Barker-Henderson first-order perturbation theory\cite{BarkerHend,HansenMcdonald},%
\begin{equation}
F\left( \overline{\rho }\right) =F_{HS}\left( \overline{\rho };d_{HS}\right)
+\frac{1}{2V}\overline{\rho }^{2}\int \Theta \left( r_{12}-r_{0}\right)
v\left( r_{12}\right) g_{HS}\left( r_{12};d_{HS}\right) d\mathbf{r}_{1}d%
\mathbf{r}_{2},
\end{equation}%
where $g_{HS}\left( r_{12};d_{HS}\right) $ is the hard-sphere pair
distribution function in the fluid phase. 

\section{Results}

\subsection{Surface Tension}
\begin{figure*}[tbp]
\begin{center}
\resizebox{12cm}{!}{
{\includegraphics[height = 12cm, angle=-90]{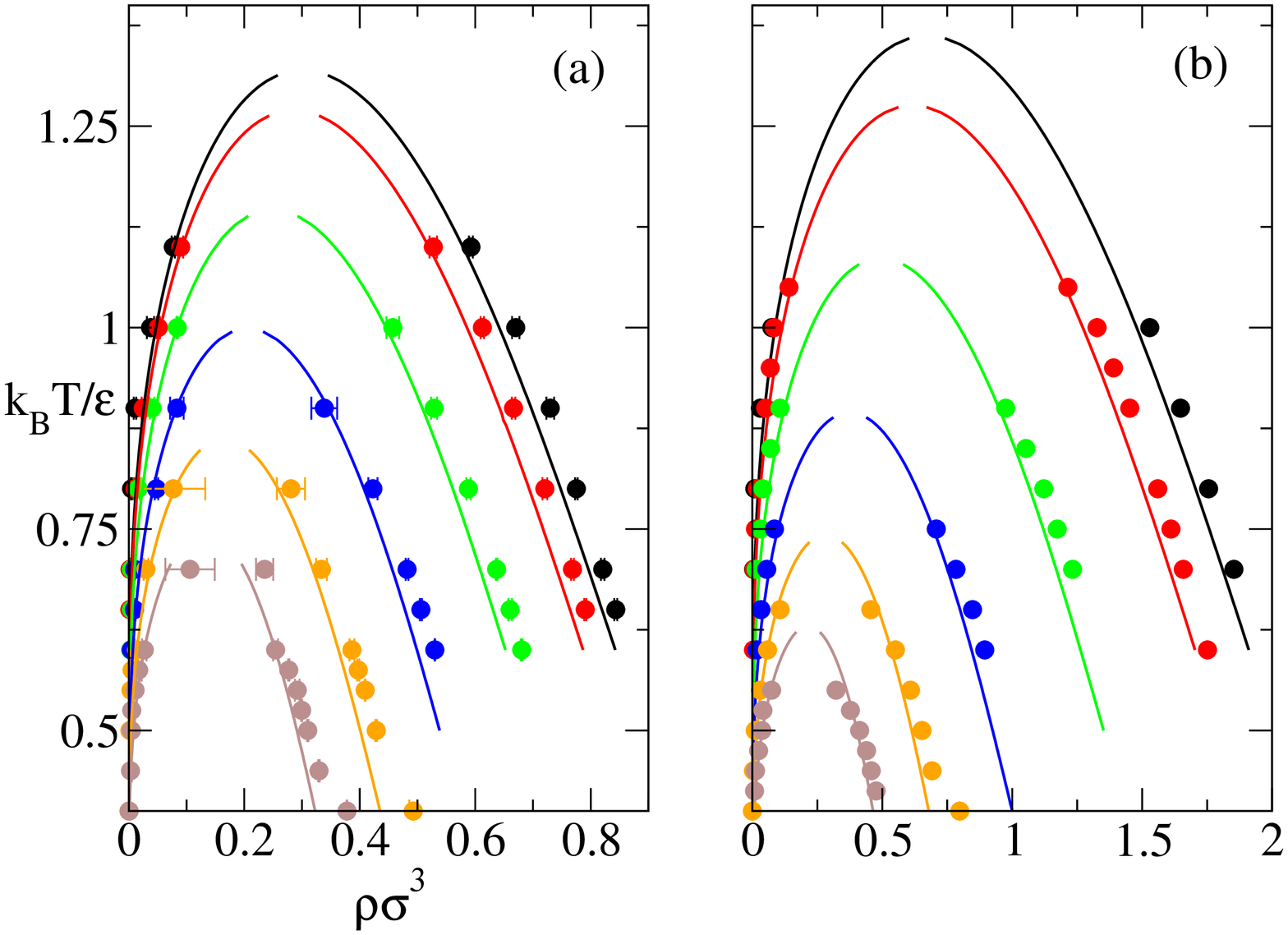}}}
\end{center}
\caption{(Color online) The phase diagrams for $\alpha = 1$, panel (a), and for $\alpha = 5$, panel (b),
as determined by simulation and theory for different values of the hard-core radius, $R=\delta \sigma$. The curves, from
top to bottom, are for  $\revision{\delta} = 0,0.2,0.4,0.6,0.8$ and $1.0$, respectively. }
\label{fig1}
\end{figure*}

\begin{figure*}[tbp]
\begin{center}
\resizebox{12cm}{!}{
{\includegraphics[height = 12cm, angle=-90]{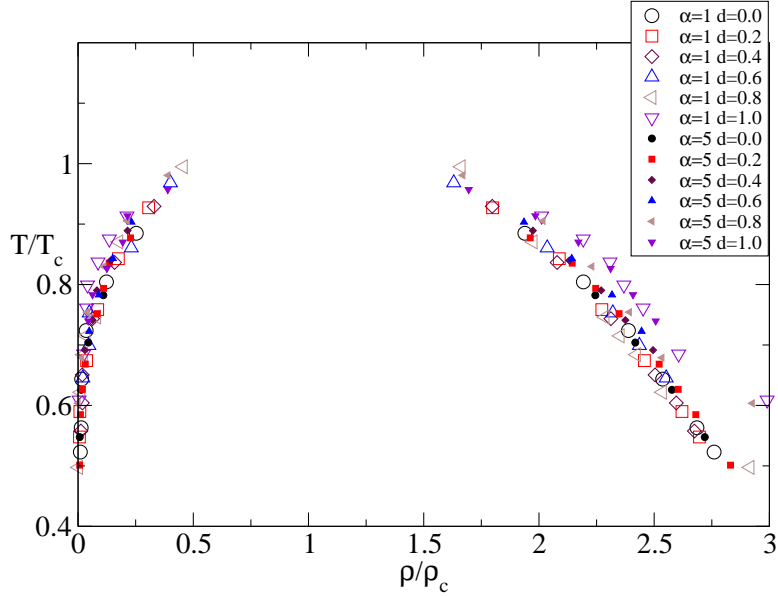}}}
\end{center}
\caption{(Color online) The same as \ref{fig1} with the density and temperature scaled to the critical density and temperature for each potential.}
\label{fig2}
\end{figure*}

\ref{fig1} shows the phase diagrams obtained from the simulations as
described above and as predicted using the thermodynamic perturbation theory
for the cases of $\alpha =1$ and $\alpha =5$. Away from the critical point,
the agreement between theory and simulation is satisfactory over a wide
range of hard-core radii. As is usual for a mean-field theory, the
agreement near the critical point is not expected to be very good and this
region has therefore not been studied. The critical density and temperature for each system was estimated following the procedure in Ref.\cite{PGLutsko} and
are given  in \ref{tab1} and \ref{tab2}. The coexistence curves are shown in 
\ref{fig2} with the density and temperature scaled
to the critical density and critical temperature respectively. Although the generalized potential involves two length scales, $\sigma$ controlling the position of the minimum and the hard-sphere radius $\delta \sigma$, the coexistence curves show typical corresponding-states as has been seen for other multi-length scale potentials\cite{PGLutsko}.

\begin{figure*}[tbp]
\begin{center}
\resizebox{12cm}{!}{
{\includegraphics[height = 12cm, angle=-90]{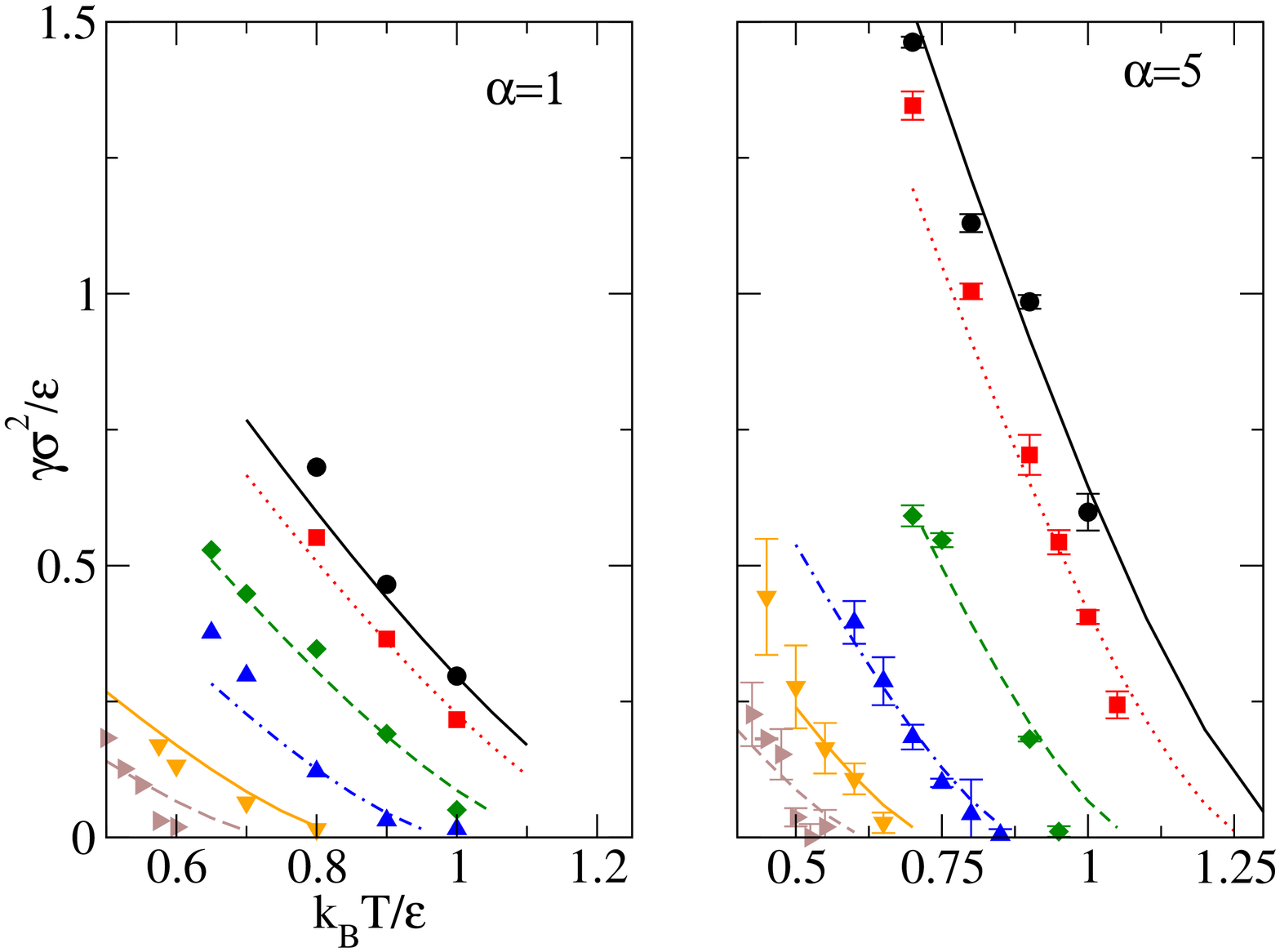}}}
\end{center}
\caption{(Color online) The surface tension as a function of temperature for  $\alpha = 1$, panel (a), and for $\alpha = 5$, panel (b),
as determined by simulation and theory for different values of the hard-core radius, $R = \delta \sigma$. The curves, from
top to bottom, are for  $\delta = 0,0.2,0.4,0.6,0.8$ and $1.0$, respectively. }
\label{fig3}
\end{figure*}

A comparison between the surface tension as determined by simulation and
from DFT calculations is shown in \ref{fig3}. Increasing the size of the molecule, i.e. the excluded volume parameter $\delta$, at fixed $\alpha$ leads to lower surface tension: since the surface tension scales, roughly, with the critical temperature, see \ref{tab1}, this trend is attributable to the decrease of the critical temperature with increasing $\delta$ which, in turn, is due to the increase with $\delta$ of  the range of repulsion (which extends from $r=0$ to $r=\delta \sigma + \sqrt{\delta^2+\left(\frac{2}{\alpha}\right)^{1/3}}$) compared to the range of the attractive part of the potential (and taking into account that the depth of the potential is fixed).  On the other hand, for fixed $\delta$, increasing $\alpha$ has little effect for $\delta = 1$ but leads to a significant increase in the surface tension for $\delta = 0$. In the latter case, the potential can be written as a function of the single parameter $\sigma/\alpha^{1/6}$ so that one expects the surface tension in this case to vary as $\gamma \sim \alpha^{1/3}$. (Note that in fact this scaling is somewhat spoiled by the fact that the cutoff used in the simulations was not scaled in the same way, so this should only be taken as an explanation for the general trends.) In the former case of $\delta = 1$, the effect of changing $\alpha$ is mitigated by the relatively large (and fixed) effect of the excluded volume and, of course, the fixed depth of the potential minimum.

Since the surface tension goes to zero at the critical point and, as discussed above, the perturbative equation of state is least accurate near the critical point, it is not surprising that the relative accuracy of the theoretical calculations decreases as one approaches the critical point. Away from this region, however, the  theoretical calculations are in good, nearly quantitative agreement with the simulations. Based on this comparison, the extension of the calculations to higher values of $\alpha$ seems justified. 

\ref{tab1} and \ref{tab2} also show the result of a fit to the expected form $\gamma = \gamma_0(1-T/T_c)^{1.26}$. Fitting both $\gamma_0$ and $T_c$ and comparing to the critical temperature extracted from the coexistence data gives a consistency check on the analysis. In general, the temperatures derived by both methods are in reasonable agreement. The errors in the critical density are too large to permit us to make a definitive statement concerning whether or not the surface tensions obey a law of corresponding states.

\begin{table}[tbp]
\caption{Fits to surface tension: $\gamma(T) = \gamma_0 T_c (1-T/T_c)^{1.26}$ for $\alpha = 1$.The last two columns give the critical properties estimated from the coexistence data.}
\label{tab1}%
%\begin{ruledtabular}
\begin{tabular}{ccccc}
d & $T_c$ & $\gamma_0$ & $T_c$ (coex) & $\rho_c$ (coex)\\ \hline
0.0 & 1.21 & 2.15  & 1.24 & 0.31\\
0.2 & 1.15 & 2.30  & 1.19 & 0.29\\
0.2 & 1.18 & 1.95  & 1.19 & 0.29\\
0.4 & 1.09 & 1.55  & 1.08 & 0.25\\
0.4 & 1.06 & 1.95  & 1.08 & 0.25\\
0.6 & 0.92 & 1.87  & 0.93 & 0.21\\
0.8 & 0.80 & 1.68  & 0.80 & 0.17\\
1.0 & 0.61 & 2.52  & 0.66 & 0.15\\
\end{tabular}
%\end{ruledtabular}
\end{table}

\begin{table}[tbp]
\caption{Fits to surface tension: $\gamma(T) = \gamma_0 T_c (1-T/T_c)^{1.26}$ for $\alpha = 5$. The last two columns give the critical properties estimated from the coexistence data.}
\label{tab2}%
%\begin{ruledtabular}
\begin{tabular}{ccccc}
d & $T_c$ & $\gamma_0$ & $T_c$ (coex) & $\rho_c $ (coex)\\ \hline
0.0 & 1.33 & 2.83  & 1.28 & 0.68 \\
0.2 & 1.18 & 3.53  & 1.20 & 0.62 \\
0.4 & 0.99 & 2.96  & 1.01 & 0.49 \\
0.6 & 0.85 & 2.09  & 0.83 & 0.37 \\
0.8 & 0.66 & 2.63  & 0.66 & 0.27 \\
1.0 & 0.54 & 3.06  & 0.57 & 0.19 \\
\end{tabular}
%\end{ruledtabular}
\end{table}

\subsection{Nucleation of globular proteins}
\begin{figure*}[tbp]
\begin{center}
\resizebox{12cm}{!}{
{\includegraphics[height = 12cm, angle=-90]{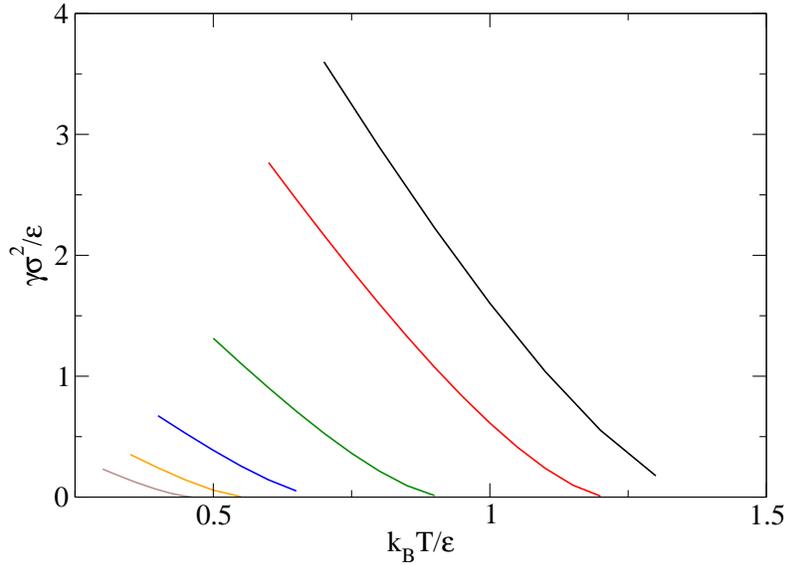}}}
\end{center}
\caption{(Color online) The surface tension as a function of temperature for  $\alpha = 50$ as calculated using DFT. The curves, from
top to bottom, are for  $\delta = 0,0.2,0.4,0.6,0.8$ and $1.0$, respectively. }
\label{fig5}
\end{figure*}

Globular proteins are modeled using $\alpha = 50$ and $\delta = 1$ so that the liquid is metastable as shown in \ref{fig1a}. As discussed above, this makes simulation extremely difficult so that we only present the results of calculations using DFT. \revision{Indeed, the fact that the metastable phase immediately tends to nucleate the solid phase in simulation is one of the main reasons for focusing on the nucleation of the metastable phase as this appears to be the rate-limiting step.} The agreement found above between DFT and simulation for lower values of $\alpha$ suggests that these results should be reasonable quantitative estimates. \ref{fig5} shows the surface tension as a function of temperature for different values of the hard core radius. The trends noted earlier are repeated with the surface tension for $\delta = 0$ being approximately $50^{1/3} \sim 3.7$ times than for $\alpha=1$ while the values of $\delta = 1$ are comparable to those for lower values of $\alpha$.  

The liquid-vapor transition is of particular interest as the metastable liquid phase is thought to play a key role in the process of precipitation of solid protein crystals from solution\cite{VekilovCGDReview2004,tWF,LutskoNicolis}. In classical nucleation theory (CNT) a liquid droplet is treated in the capillary approximation so that a droplet of radius $R$ has excess free energy 
\begin{equation}
\Delta \Omega = \frac{4\pi}{3}R^{3}\left( \omega(\rho_l) - \omega(\rho_v)\right) + 4\pi R^{2} \gamma
\end{equation}
where $\omega(\rho) = f(\rho)-\mu \rho$, $f(\rho$)  is the bulk Helmholtz free energy per unit volume  of the fluid at density $\rho$, $\mu$ is the chemical potential, $\gamma$ is the liquid-vapor surface tension at coexistence and $\rho_l$ and $\rho_v$ are the densities of the coexisting liquid and vapor respectively. In this model, the excess number of molecules in the droplet is
\begin{equation}
\Delta n = \frac{4\pi}{3}R^{3}(\rho_l - \rho_v).
\end{equation}
The free energy has a maximum at the critical radius, $R_c = 2\gamma/\Delta\omega$, with
$\Delta\omega=\omega(\rho_v)-\omega(\rho_l)$,  and a maximum, defining the barrier for nucleation, of $\Delta \Omega_{max} = \frac{16\pi\gamma^3}{3\Delta\omega^2}$. Thus, from these simple considerations and the previously noted trends in the surface tension, we conclude that if all of the energy scales are proportional to $T_{c}$, then the nucleation barrier will decrease with increasing molecular size ($\delta$) and will vary weakly with the range of the potential, $\alpha$.

\begin{figure*}[tbp]
\begin{center}
\resizebox{12cm}{!}{
{\includegraphics[height = 12cm, angle=-90]{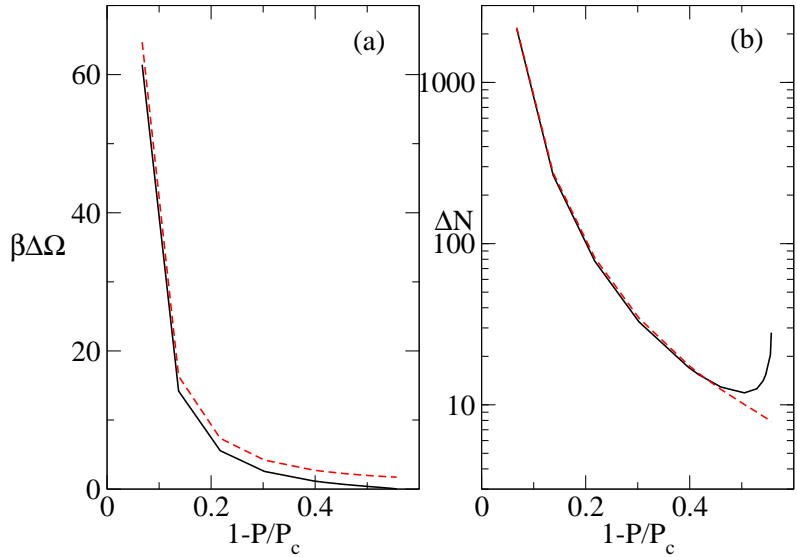}}}
\end{center}
\caption{(Color online) Panel (a) shows the free energy barrier for nucleation of dense liquid from the low-density phase as a function of supersaturation. Panel (b) shows the size of the critical cluster. In both cases, the full lines are from the DFT calculations and the broken lines are the prediction of CNT.} 
\label{fig6}
\end{figure*}

\begin{figure*}[tbp]
\begin{center}
\resizebox{12cm}{!}{
{\includegraphics[height = 12cm, angle=-90]{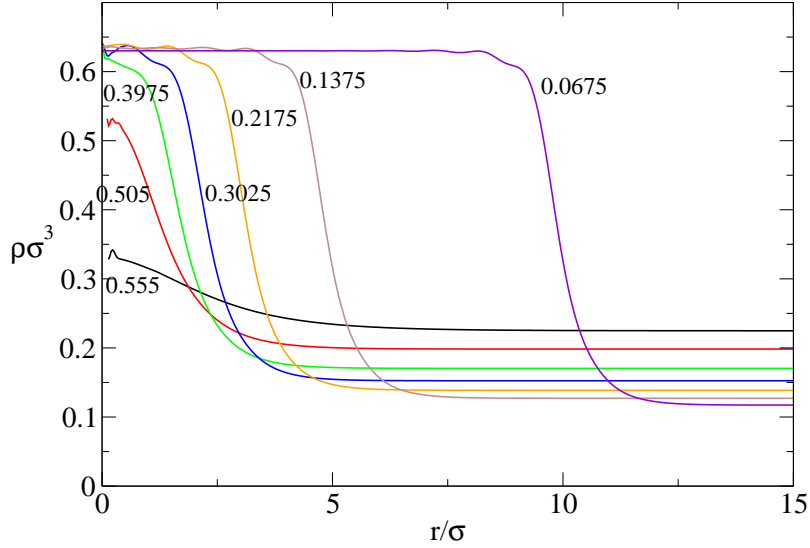}}}
\end{center}
\caption{(Color online) The density distribution in the critical clusters for various values of the supersaturation.}
\label{fig7}
\end{figure*}

\begin{figure*}[tbp]
\begin{center}
\resizebox{12cm}{!}{
{\includegraphics[height = 12cm, angle=-90]{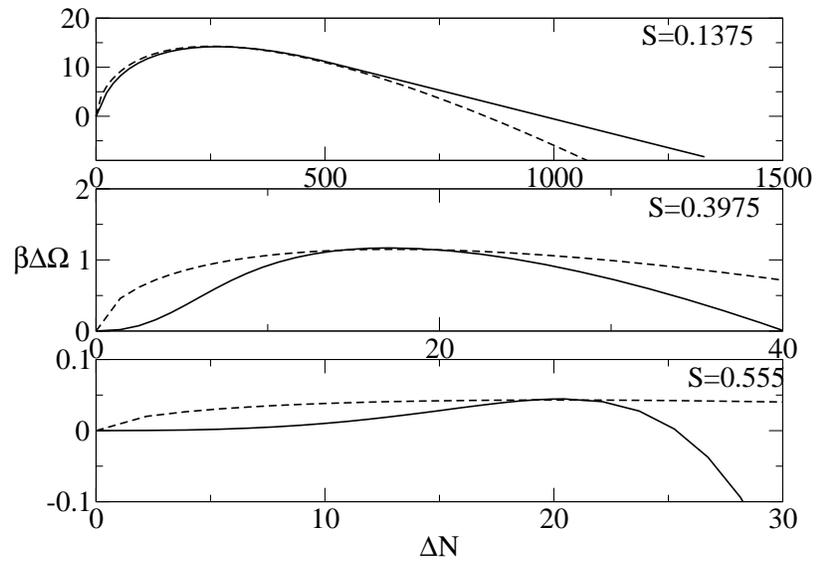}}}
\end{center}
\caption{The excess free energy as a function of cluster size at three different values of supersaturation. The broken lines are a fit to the CNT expression $\Delta \Omega = -\frac{2 \Delta \Omega_*}{\Delta N_*}\Delta N + \frac{3 \Delta \Omega}{\Delta N_*^{2/3}}\Delta N^{2/3}$ where the starred-quantities refer to the critical nucleus.}
\label{fig8}
\end{figure*}

To investigate the liquid-vapor transition in more detail, we have determined the nucleation pathway (specifically, the minimum free energy path or MFEP) for nucleation of liquid droplets from the DFT to compare to the predictions of CNT for $T=0.4\epsilon$ or $T/T_c = 0.84$. The method of calculation is the same as described in detail in Refs.\cite{LutskoEPL,Lutsko_JCP_2008_2}. The goal is to identify a path between the initial state (pure vapor) and the final state (pure liquid) which is minimal with respect to variations perpendicular to the path (for a detailed explanation, see e.g.  \cite{Wales}). \ref{fig6} shows the free energy barrier and the size of the critical cluster as a function of supersaturation, $S=\frac{P_{coex}-P}{P_{coex}}$ where the chemical potential at coexistence is $\beta \mu_{coex} = -2.354$ and the densities of the coexisting phases are $\rho_v =0.109$ and $\rho_l = 0.626$. (Note that although the actual control variable in our calculations in the grand canonical ensemble is chemical potential, we use the more familiar definition of supersaturation in terms of the pressure.) \ref{fig6} shows that classical nucleation theory is in agreement with DFT for small supersaturations but for large supersaturation, the nucleation barrier calculated with DFT becomes very small, approaching zero as the vapor density approaches the spinodal, whereas CNT predicts a finite nucleation barrier at all densities. The figure also shows that the excess number of molecules in the critical cluster is also well described by CNT except at large supersaturations.
\ref{fig7}, showing the critical clusters at different supersaturations,  illustrates the reason for the failure of CNT at large supersaturation. It shows that large clusters are indeed well described with the capillary model, having very narrow interfaces. However, at large supersaturation, the critical clusters are very small with most molecules affected by the interface and with central densities far below that of the bulk liquid. As a consequence, CNT fails to capture the very small barriers for nucleation. This final point is illustrated in \ref{fig8} which shows the nucleation pathway, i.e. the excess free energy as a function of cluster size,  for three different supersaturations. At the smallest supersaturation, the barrier is well described by CNT. However, at large supersaturation, the shape of the nucleation barrier varies markedly from that assumed in CNT. In particular, there is virtually no free energy penalty for the formation of small clusters.

\begin{figure*}[tbp]
\begin{center}
\resizebox{12cm}{!}{
{\includegraphics[height = 12cm, angle=-90]{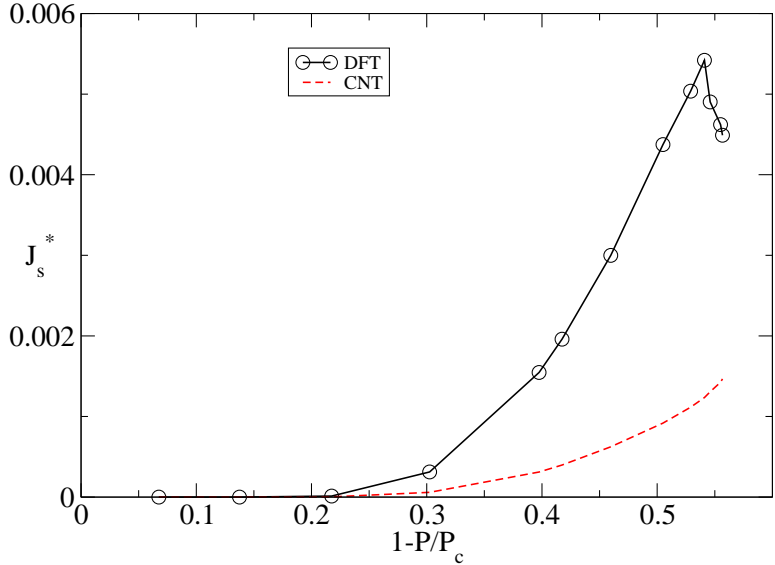}}}
\end{center}
\caption{(Color online) Dimensionless nucleation rate as a function of the supersaturation. The scaled nucleation rate is $J_{s}^{*}=J\sigma^3/f_{0}$.}
\label{fig9}
\end{figure*}

Given that we have access to the nucleation pathway, and not just the critical cluster, it is possible to directly evaluate the nucleation rate. Under the assumption of stationary nucleation and of treating the number of molecules as a continuous variable, the nucleation rate is given by the otherwise exact expression
\begin{equation}
J_{s}=\left( \int_1^\infty \frac{dn}{f(n)C(n)} \right)^{-1}
\end{equation} 
where $f(n)$ is the monomer attachment rate for a cluster of size $n$ and $C(n)= C_{0} exp(-\Omega(n)/k_{B}T)$ is the equilibrium distribution of cluster sizes.\cite{Zeldovich1, Zeldovich2,Kashchiev}. The monomer attachment rate is assumed to be proportional to the surface area of a cluster and to the gas pressure, $f(n) = f_0(P/k_BT)n^{2/3}$ with $f_0 = \gamma c v_{0}^{2/3}(\frac{k_BT}{2\pi m_{0}})^{1/2}$ where $\gamma$ is the sticking probability for a monomer that collides with a cluster of size $n$ (assumed here to be independent of $n$), $c$ is a geometric factor and $m_{0}$ and $v_0$ are  the mass and volume of a molecule so that the area of the cluster is $c v_{0}^{2/3}n^{2/3}$\cite{Kashchiev}. \ref{fig9} shows the calculated nucleation rate as a function of the supersaturation. The nucleation rate estimated from DFT is significantly higher than that estimated from CNT, partly due to the exponential amplification of the differences in barrier height seen in \ref{fig6} and partly due to the differences in shape of the barrier. 

\ref{fig6} and \ref{fig9} both show non-monotonic behavior near the spinodal. In the later case, the implication is that there is an optimal supersaturation that gives the maximal nucleation rate. However, this is likely to be an artifact of the calculations. As discussed in detail by Wilemski and Li\cite{Wilemski}, mean field models such as that used here, and simpler models such as the square gradient model, predict a divergent critical nucleus size at the spinodal due to the the unrealistic mean-field equation of state. In any case, since the nucleation barrier is on the order of $k_{B}T$ for  $S \sim 0.4$, it is likely that the process of nucleation at higher supersaturations shares features of both nucleation and spinodal decomposition sometimes termed ``spinodal nucleation''\cite{Binder, Wilemski} in which case the nucleation characteristics (rate and critical nucleus) would not be described by the simple theory given here.  There are, therefore, two issues affecting the interpretation of these results. The first is the validity of mean field theory, which definitely breaks down near the spinodal\cite{Binder, Wilemski} and the second is that even if the maximum occurred in the region of validity of the theory, it would be masked by density fluctuations that would cause small volumes to become unstable with respect to spinodal decomposition. 

\section{Conclusions}
In this paper, we have constructed a pair potential that allows us to move smoothly from a simple fluid (i.e. Lennard-Jones interaction) to the short-ranged, ``hard Lennard-Jones'' used as a model for globular protein interactions. The potential depends on two parameters: the size of the molecule, characterized by the excluded volume parameter $\delta$, and the range of the potential, controlled by the dimensionless parameter $\alpha$. We compared the liquid-vapor surface tension as determined from Monte Carlo simulations to DFT calculations for the case $\alpha =1,5$ and molecular radius ranging from zero to one (in Lennard-Jones units). Our results indicate that the combination of thermodynamic perturbation theory and the MC-VDW DFT model give a good quantitative description of the liquid-vapor equation of state and surface tension over a wide range of temperatures. Significant differences do appear as expected near the critical point since the mean-field equation of state is not accurate in this region. Our primary conclusion from these calculations is that it is the increase in excluded volume rather than the decrease in range of the potential that causes a dramatic decrease in the surface tension. In fact, from \ref{fig3} and \ref{fig5}, one sees that while the surface tension for zero hard-core radius is strongly affected by the range of the potential, the surface tension at the largest excluded volume is relatively insensitive to the range. 

\revision{For fixed $\delta$, changing $\alpha$ changes the range of the potential. However, the model potential can in fact be rewritten in terms of a single dimensioness parameter, $\alpha \delta^6$ so that in absence of a cutoff, changing $\alpha$ and changing $\delta$ are in some sense equivalent. The actual physical relevance of the two parameters is that $\delta$ controls the size of the exluded volume while $\alpha$ controls the strength of the attractive interaction. Hence, one way to phrase our results is that changing the excluded volume has much more effect on physical properties, such as surface tension, than does changing the strength of the attraction. These results suggest that increasing the excluded volume of a molecule leads to a decrease in surface tension and hence of the barrier to nucleation of the dense phase. Conformational changes can in fact be affected in some cases by changes in pH\cite{D1,D3,D4} and by light\cite{D2} thus suggesting a means for taking advantage of this phenomena to control crystallization rates in such systems.}

For the case $\alpha=50$, we were not able to simulate the liquid-vapor interface due to the strong instability of the system to crystallization. Our theoretical calculations indicate similar behavior in the surface tension as found for smaller values of $\alpha$. We have also calculated the nucleation pathway and thereby determined the barrier and rate of nucleation of liquid droplets from the vapor. We found quite low barriers, less than $100k_BT$ even at relatively low supersaturation. Away from the spinodal, Classical Nucleation Theory gives a good description of the barrier height and the size of the critical nucleus. At higher supersaturations, as the spinodal is approached, the nucleation pathway calculated from DFT differs significantly from that predicted by CNT. DFT also predicts a vanishing nucleation barrier at the spinodal and non-monotonic behavior of the size of the critical nucleus and the nucleation rate near the spinodal. The shape of the barriers suggests that the excess free energy of small, sub-critical clusters is very small indicating that they might have relatively long lifetimes. However, the details of this picture become increasingly uncertain near the spinodal due to the limitations of mean field theory and in any case may not be observable as they occur in the region of ``spinodal nucleation''.

%\begin{acknowledgments}
\begin{acknowledgement}
\revision{Our work has benefited from several discussions with Dominique Maes at the Flanders Interuniversity Institute for Biotechnology (VIB),Vrije Universiteit Brussel.} This work was supported by the European Space Agency under contract number
ESA AO-2004-070 and by the projet ARCHIMEDES of the Communaut\'e Fran\c
caise de Belgique (ARC 2004-09).
%\end{acknowledgments}
\end{acknowledgement}

\bibliography{protein}

\end{document}